\documentclass[twocolumn,floatfix,showpacs,superscriptaddress]{revtex4}
\usepackage{graphicx}
\usepackage{amsmath}
\usepackage{amssymb}
\usepackage{bm}
\usepackage{esint} % used for the symbol \fint

\begin{document}

\title{Fermion-parity anomaly of the critical supercurrent in the quantum spin-Hall effect}
\author{C. W. J. Beenakker}
\affiliation{Instituut-Lorentz, Universiteit Leiden, P.O. Box 9506, 2300 RA Leiden, The Netherlands}
\author{D. I. Pikulin}
\affiliation{Instituut-Lorentz, Universiteit Leiden, P.O. Box 9506, 2300 RA Leiden, The Netherlands}
\author{T. Hyart}
\affiliation{Instituut-Lorentz, Universiteit Leiden, P.O. Box 9506, 2300 RA Leiden, The Netherlands}
\author{H. Schomerus}
\affiliation{Department of Physics, Lancaster University, Lancaster, LA1 4YB, United Kingdom}
\author{J. P. Dahlhaus}
\affiliation{Instituut-Lorentz, Universiteit Leiden, P.O. Box 9506, 2300 RA Leiden, The Netherlands}

\date{October 2012}
\begin{abstract}
The helical edge state of a quantum spin-Hall insulator can carry a supercurrent in equilibrium between two superconducting electrodes (separation $L$, coherence length $\xi$). We calculate the maximum (critical) current $I_{\rm c}$ that can flow without dissipation along a single edge, going beyond the short-junction restriction $L\ll\xi$ of earlier work, and find a dependence on the fermion parity of the ground state when $L$ becomes larger than $\xi$. Fermion-parity conservation doubles the critical current in the low-temperature, long-junction limit, while for a short junction $I_{\rm c}$ is the same with or without parity constraints. This provides a phase-insensitive, {\sc dc} signature of the $4\pi$-periodic Josephson effect.
\end{abstract}
\maketitle

The quantum Hall effect and quantum spin-Hall effect both refer to a two-dimensional semiconductor with an insulating bulk and a conducting edge, and both exhibit a quantized electrical conductance between two metal electrodes. If the electrodes are superconducting, a current can flow in equilibrium, induced by a magnetic flux without any applied voltage. In the quantum Hall effect, the edge states are chiral (propagating in a single direction only) and two opposite edges are needed to carry a supercurrent \cite{Ma93,Ost11,Sto11}. Graphene is an ideal system to study this interplay of the Josephson effect and the quantum Hall effect in a strong magnetic field \cite{Ric12,Pop12,Kom12}.

The interplay of the Josephson effect and the quantum spin-Hall effect, in zero magnetic field, has not yet been demonstrated experimentally but promises to be strikingly different \cite{Fu09}. The quantum spin-Hall insulator has helical edge states (propagating in both directions) that can carry a supercurrent along a single edge. The edge state couples a pair of Majorana zero-modes, allowing for the transmission of unpaired electrons with $h/e$ rather than $h/2e$ periodic dependence on the magnetic flux \cite{Kit01,Kwo03}.

An $h/e$ flux periodicity corresponds to a $4\pi$-periodicity in terms of the superconducting phase difference $\phi$, which means that the current-phase relationship has two branches $I_{\pm}(\phi)$ and the system switches from one branch to the other when $\phi$ is advanced by $2\pi$ at fixed total number ${\cal N}$ of electrons in the system. This is referred to as a fermion-parity anomaly, because the two branches have different parity $\sigma=\pm$ of the number of electrons in the superconducting ground state \cite{Kit01}.

Josephson junctions come in two types \cite{Gol04}, depending on whether the separation $L$ of the superconducting electrodes is small or large compared to the coherence length $\xi=\hbar v/\Delta$, or equivalently, whether the superconducting gap $\Delta$ is small or large compared to the Thouless energy $E_{\rm T}=\hbar v/L$. Existing literature \cite{Fu09,Kit01,Kwo03,Lut10,Law11,Ios11,Nog12,Bad11,San12,Dom12,Pik11} has focused on the short-junction regime $L\ll\xi$. The supercurrent is then determined entirely by the phase dependence of a small number of Andreev levels in the gap, just one per transverse mode. The phase dependence of the continuous spectrum above the gap can be neglected. As the ratio $L/\xi$ increases, the Andreev levels proliferate and also the continuous spectrum starts to contribute to the supercurrent. Since $\sigma$ is switched by changing the occupation of a single level, one might wonder whether a significant parity dependence remains in the long-junction regime.

\begin{figure*}[tb]
\centerline{\includegraphics[width=0.9\linewidth]{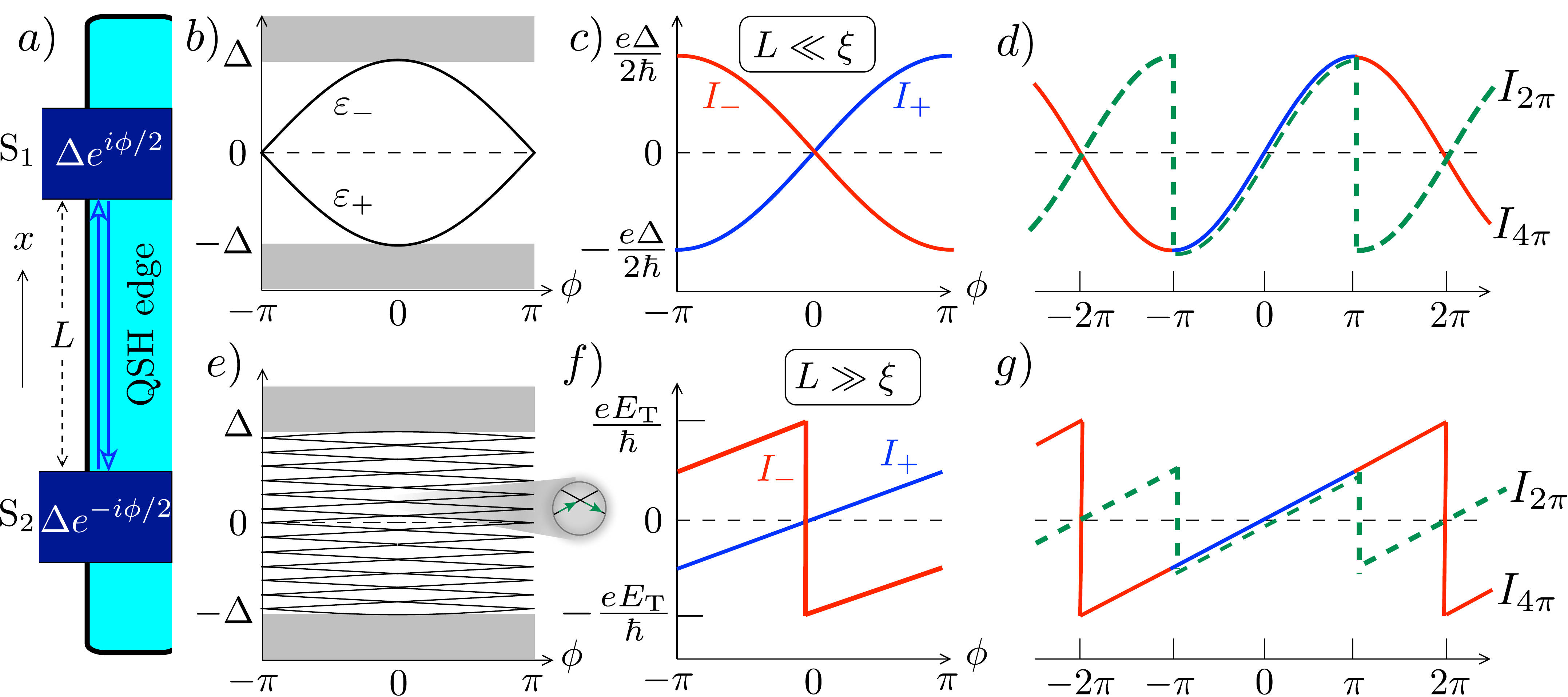}}
\caption{Phase-dependent excitation spectrum of a Josephson junction along a quantum spin-Hall  (QSH) edge (left panels) and corresponding zero-temperature supercurrent (right panels). The supercurrent $I_{4\pi}$ is $4\pi$-periodic, with two branches $I_{+}$ (blue solid), $I_{-}$ (red solid) distinguished by the ground-state fermion parity and with a parity switch at $\phi=\pm\pi$. The top row shows the short-junction limit of Ref.\ \onlinecite{Fu09}, the bottom row the long-junction limit calculated here. (The jump in $I_{-}$ at $\phi=0$ occurs because of the change in slope indicated by the green arrows in the magnified central part of the spectrum.) The $2\pi$-periodic supercurrent $I_{2\pi}$ without parity constraints is also shown (green dashed). The critical current is the same for $I_{4\pi}$ and $I_{2\pi}$ in the short junction, but different by a factor of two in the long junction.
}
\label{fig_spectrum}
\end{figure*}

Remarkably enough, the parity dependence becomes even stronger. While in a short junction the two branches $I_{+}(\phi)=-I_{-}(\phi)$ differ only in sign, we find that in a long junction they differ both in sign and in magnitude. In particular, the largest current that can flow without dissipation is twice as large for $I_{-}$ than it is for $I_{+}$. The difference is illustrated in Fig.\ \ref{fig_spectrum}, in the zero-temperature limit. The basic physics can be explained in simple terms, as we will do first, and then we will present a complete theory for finite temperature and for arbitrary ratio $L/\xi$.

We set the stage by summarizing the findings of Fu and Kane \cite{Fu09} in the short-junction regime. The spectrum of the Bogoliubov-De Gennes Hamiltonian $H_{\rm BdG}$ is a $\pm\varepsilon$ symmetric combination of a discrete spectrum for $|\varepsilon|<\Delta$ and a continuous spectrum for $|\varepsilon|>\Delta$. Since backscattering along the quantum spin-Hall edge is forbidden by time-reversal symmetry \cite{Kon08}, this is a ballistic single-channel Josephson junction. In the limit $L/\xi\rightarrow 0$ the discrete spectrum consists of a pair of levels at $\varepsilon_{\pm}=\mp\Delta|\cos(\phi/2)|$, while the continuous spectrum is $\phi$-independent \cite{Bee91a}. Quite generally, an eigenvalue $\varepsilon(\phi)$ of $H_{\rm BdG}$ contributes to the supercurrent an amount
\begin{equation}
I(\phi)=\frac{ge}{\hbar}\frac{d}{d\phi}\varepsilon(\phi),\label{Iphirelation}
\end{equation}
with $g$ a factor that counts spin and other degeneracies \cite{note1}. There is no spin degeneracy at the quantum spin-Hall edge (since spin is tied to the direction of motion), so $g=1$ and the level $\varepsilon_{\pm}$ contributes a supercurrent \cite{Fu09}
\begin{equation}
I_{\pm}(\phi)=\pm\frac{e\Delta}{2\hbar}\sin(\phi/2),\;\;|\phi|<\pi.\label{Iphipm}
\end{equation}

To discuss the fermion-parity anomaly we assume, for definiteness, that the total number ${\cal N}$ of electrons in the system is even. (A different choice amounts to a $2\pi$ phase shift, or equivalently, an interchange of $I_{+}$ and $I_{-}$.) The ground-state fermion parity $\sigma$ is even for $\phi=0$ and switches to odd when $\phi$ crosses $\pi$. Since ${\cal N}$ is fixed, this topological phase transition must be accompanied by a switch between even and odd number of quasiparticle excitations. At zero temperature only the two levels $\varepsilon_{\pm}$ closest to the Fermi level ($\varepsilon=0$) play a role, and the parity switch of $\sigma$ means that a quasiparticle is transferred from $\varepsilon_{+}<0$ to $\varepsilon_{-}>0$. It cannot relax back from $\varepsilon_{-}$ to $\varepsilon_{+}$ at fixed parity of ${\cal N}$.

The resulting current-phase relationship can be represented by a switch between $2\pi$-periodic branches $I_{\pm}(\phi)$ (reduced zone scheme), or equivalently as a $4\pi$-periodic function $I_{4\pi}(\phi)$ (extended zone scheme).  Both representations are shown in Fig.\ \ref{fig_spectrum}, upper panels. We also include the $2\pi$-periodic current $I_{2\pi}$ that results if the system can relax to its lowest energy state without constraints on the parity of ${\cal N}$.

So much for the short-junction limit. An elementary discussion of the long-junction regime (to be made rigorous in just a moment) goes as follows. For $L\gg\xi$ we may assume \cite{Ish71,Bar72,Svi73} a local linear relation between the current density $I$ and the phase gradient $\phi/L\ll 1/\xi$, of the form $I={\rm constant}\times ev\phi/L$. The linear increase of $I_{-}$ is interrupted at $\phi=0$ by a discontinuity $\Delta I_{-}=2ev/L$. Half of it results from the jump in the slope of the lowest occupied positive energy level $\varepsilon=(\pi-|\phi|)\hbar v/2L$ (green arrows in Fig.\ \ref{fig_spectrum}e). The jump in the slope of the highest occupied negative energy level contributes the other half. In the extended zone scheme, the resulting supercurrent $I_{4\pi}$ is a $4\pi$-periodic sawtooth with a slope $\Delta I_{-}/4\pi=eE_{\rm T}/2\pi\hbar$.

The corresponding parity-dependent supercurrents in the reduced zone scheme are
\begin{equation}
I_{+}=\frac{eE_{\rm T}}{2\pi\hbar}\phi,\;\;I_{-}=\frac{eE_{\rm T}}{2\pi\hbar}(\phi-2\pi\,{\rm sign}\,\phi),\;\;|\phi|<\pi.\label{Iphipmlong}
\end{equation}
The $4\pi$-periodic supercurrent $I_{4\pi}$ switches from $I_{+}$ to $I_{-}$ at $\phi=\pi$, while $I_{2\pi}$ remains in the branch $I_{+}$ by compensating the switch in ground-state fermion parity $\sigma$ by a switch in the parity of the electron number ${\cal N}$. These are the curves plotted in Fig.\ \ref{fig_spectrum} (lower panels).

Looking at the upper panels, one might have expected the sinusoidal current-phase relationship of $I_{4\pi}$ for a short junction to evolve into a triangular profile for a long junction, remaining symmetric around $\phi=\pi$. This would produce a cusp at the topological phase transition, which is avoided by the sawtooth profile --- at the expense of a discontinuity at $\phi=2\pi$.

The maximal supercurrent is reached near $\phi=2\pi$ for $I_{4\pi}$ (with parity constraint) and near $\phi=\pi$ for $I_{2\pi}$ (without parity constraint). There is a factor of two difference in magnitude of these critical currents in a long junction,
\begin{equation}
I_{4\pi,{\rm c}}=eE_{\rm T}/\hbar,\;\;I_{2\pi,{\rm c}}=eE_{\rm T}/2\hbar.\label{IcET}
\end{equation}
In contrast, for a short junction both are the same (equal to $e\Delta/2\hbar$).

To determine the crossover from the short-junction limit \eqref{Iphipm} to the long-junction limit \eqref{Iphipmlong}, including the temperature dependence, we adapt the scattering theory of the Josephson effect \cite{Bee91b} to include the fermion parity constraints. Input is the scattering matrix $s_{0}$ of electrons in the normal region and the Andreev reflection matrix $r_{\rm A}$ at the normal-superconductor interfaces. These take a particularly simple $2\times 2$ form at the quantum spin-Hall edge, but our general formulas are applicable also to \textit{multi-channel} topological superconductors.

The parity-dependent partition function is \cite{Law11,Ios11,Nog12,Tuo92}
\begin{align}
Z_{\pm}&
=\tfrac{1}{2}\left(\prod_{\varepsilon>0}e^{\beta\varepsilon/2}\right)\left(\prod_{\varepsilon>0}(1+e^{-\beta\varepsilon})\pm\prod_{\varepsilon>0}(1-e^{-\beta\varepsilon})\right)\nonumber\\
&=\tfrac{1}{2}Z_{0}\biggl(1\pm\prod_{\varepsilon>0}\tanh(\beta\varepsilon/2)\biggr),\label{Zpmdef}
\end{align}
with $\beta=1/k_{\rm B}T$ and $Z_{0}=\prod_{\varepsilon>0}2\cosh(\beta\varepsilon/2)$ the partition function without parity constraints. From the expression for $Z_{\pm}$ one can see that the $\pm$ selects terms that contain an even ($+$) or an odd ($-$) number of quasiparticle excitation factors $e^{-\beta\varepsilon}$, as is dictated by the ground-state fermion parity.  The partition function $Z$ gives the free energy $F$ and hence the supercurrent $I$ \cite{note2},
\begin{align}
&I_{\pm}=\frac{2e}{\hbar}\frac{dF_{\pm}}{d\phi},\;\;F_{\pm}=-\beta^{-1}\ln Z_{\pm},\label{IpmZpm}\\
&I_{2\pi}\equiv I_{0}=\frac{2e}{\hbar}\frac{dF_{0}}{d\phi},\;\;F_{0}=-\beta^{-1}\ln Z_{0}.\label{I0Z0}
\end{align}

The density of states $\rho(\varepsilon)$ contains both the discrete spectrum for $|\varepsilon|<\Delta$ (a sum of delta functions at the Andreev levels) and the continuous spectrum for $|\varepsilon|>\Delta$, including also a contribution $\rho_{\rm S}$ from the superconducting electrodes. Scattering theory gives the expression \cite{Bee91b}
\begin{align}
&\rho(\varepsilon)={\rm Im}\,\frac{d}{d\varepsilon}\nu(\varepsilon+i0^{+})+\rho_{\rm S}(\varepsilon),\label{rhorhoSdef}\\
&\nu(\varepsilon)=-\pi^{-1}\ln{\rm Det}\,X(\varepsilon),\;\;X=(1-M)M^{-1/2},\label{nudef}\\
&M(\varepsilon)=r_{\rm A}^{\ast}(-\varepsilon) s_{0}^{\ast}(-\varepsilon)r_{\rm A}(\varepsilon)s_{0}(\varepsilon).
\label{rhoXrelation}
\end{align}
The factor $M^{-1/2}$ in the definition of $X$, as well as the term $\rho_{\rm S}$, give a $\phi$-independent additive contribution to $F_{0}$ without any effect on $I_{0}$, but we need to retain these terms here because they do enter into the parity constraint for $I_{\pm}$.

In the absence of parity constraints, Ref.\ \onlinecite{Bro97} gives the free energy
\begin{equation}
F_{0}=-\beta^{-1}\textstyle{\sum_{p=0}^{\infty}}\ln{\rm Det}\,X(i\omega_{p}),\label{F0result}
\end{equation}
as a sum over fermionic Matsubara frequencies $\omega_{p}=(2p+1)\pi/\beta$. A similar calculation \cite{App} gives the parity-dependence in the form
\begin{align}
F_{\sigma}={}&F_{0}-\beta^{-1}\ln\tfrac{1}{2}\biggl[1+\sigma e^{J_{\rm S}}\sqrt{{\rm Det}\,X(0)}\nonumber\\
&\quad\quad\times\exp\biggl(\sum_{p=1}^{\infty}(-1)^{p}\ln{\rm Det}\,X(i\Omega_{p}/2)\biggr)\biggr],\label{Fpmresult}\\
\sigma={}&{\rm sign}\,\bigl[{\rm Pf}\,(r_{\rm A}s_{0}-s_{0}^{\rm T}r_{\rm A}^{\rm T})({\rm Det}\,is_{0})^{-1/2}\bigr]_{\varepsilon=0},
\end{align}
with bosonic Matsubara frequencies $\Omega_{p}=2p\pi/\beta$. The ground-state fermion parity $\sigma$ is given in terms of the Pfaffian of the anti-symmetrized scattering matrix, evaluated at the Fermi energy. The sign ambiguity in the square root is resolved by fixing $\sigma=1$ at $\phi=0$.

Eq.\ \eqref{Fpmresult} contains a contribution from the superconducting electrodes,
\begin{equation}
J_{\rm S}=\int_{\Delta}^{\infty}d\varepsilon\,\rho_{\rm S}(\varepsilon)\ln\tanh(\beta\varepsilon/2),\label{JSdef}
\end{equation}
which only plays a role at temperatures $T\gtrsim \Delta/k_{\rm B}$. The factor $e^{J_{\rm S}}$ can therefore be replaced by unity in the long-junction regime, when $k_{\rm B}T\lesssim E_{\rm T}\ll\Delta$.

We now specify these general formulas for the quantum spin-Hall edge, with Hamiltonian \cite{note3}
\begin{equation}
H_{\rm BdG}=\begin{pmatrix}
vp\sigma_{z}+U(x)&\Delta^{\ast}(x)\sigma_{y}\\
\Delta(x)\sigma_{y}&vp\sigma_{z}-U(x)
\end{pmatrix}.\label{HBdGQSH}
\end{equation}
The edge runs along the $x$-axis, $p=-i\hbar\partial_{x}$ is the momentum operator, and the electrostatic potential is $U(x)$ (measured relative to the Fermi level). The pair potential $\Delta(x)$ vanishes in the normal region $|x|<L/2$. In the two superconducting regions we set $\Delta(x)=\Delta e^{\pm i\phi/2}$, with a step at $x=\pm L/2$. This socalled ``rigid boundary condition'' is justified for a single channel coupled to a bulk superconducting reservoir \cite{Gol04}.

A mode-matching calculation gives the scattering matrices
\begin{align}
&s_{0}=\begin{pmatrix}
0&e^{i\chi}\\
e^{i\chi}&0
\end{pmatrix},\;\;\chi(\varepsilon)=\chi_{0}+\varepsilon/E_{\rm T},\label{s0QSH}\\
&r_{\rm A}=\begin{pmatrix}
\alpha e^{i\phi/2}&0\\
0&-\alpha e^{-i\phi/2}
\end{pmatrix},\;\;\alpha(\varepsilon)=\sqrt{1-\frac{\varepsilon^{2}}{\Delta^{2}}}+\frac{i\varepsilon}{\Delta},\nonumber\\
&{\rm Det}\,X(\varepsilon)=2\cos\phi+\alpha^{2}e^{2i\varepsilon/E_{\rm T}}+\alpha^{-2}e^{-2i\varepsilon/E_{\rm T}}.\label{DetXresult}
\end{align}
We discuss the various terms in these expressions. The electron scattering matrix $s_{0}$ is purely off-diagonal, because of the absence of backscattering along the quantum spin-Hall edge. The transmission phase $\chi$ depends linearly on energy because of the linear dispersion. Electrostatic potential fluctuations contribute only to the energy-independent offset $\chi_{0}=-(\hbar v)^{-1}\int_{0}^{L}U\,dx$, which drops out in Eq.\ \eqref{nudef}. The Andreev reflection matrix $r_{\rm A}$ (from electron to hole) is unitary below the gap. Above the gap there is also propagation into the superconductor, so $r_{\rm A}$ is sub-unitary. The same expression \eqref{s0QSH} for $r_{\rm A}$ applies at all energies, evaluated at $\varepsilon+i0^{+}$ to avoid the branch cut of the square root.
Notice that for $\phi=0$ the Andreev reflection amplitudes from $S_{1}$ and $S_{2}$ differ by a minus sign, because of the opposite spin of counterprogating electrons in a helical edge state.

\begin{figure}[tb]
\centerline{\includegraphics[width=1\linewidth]{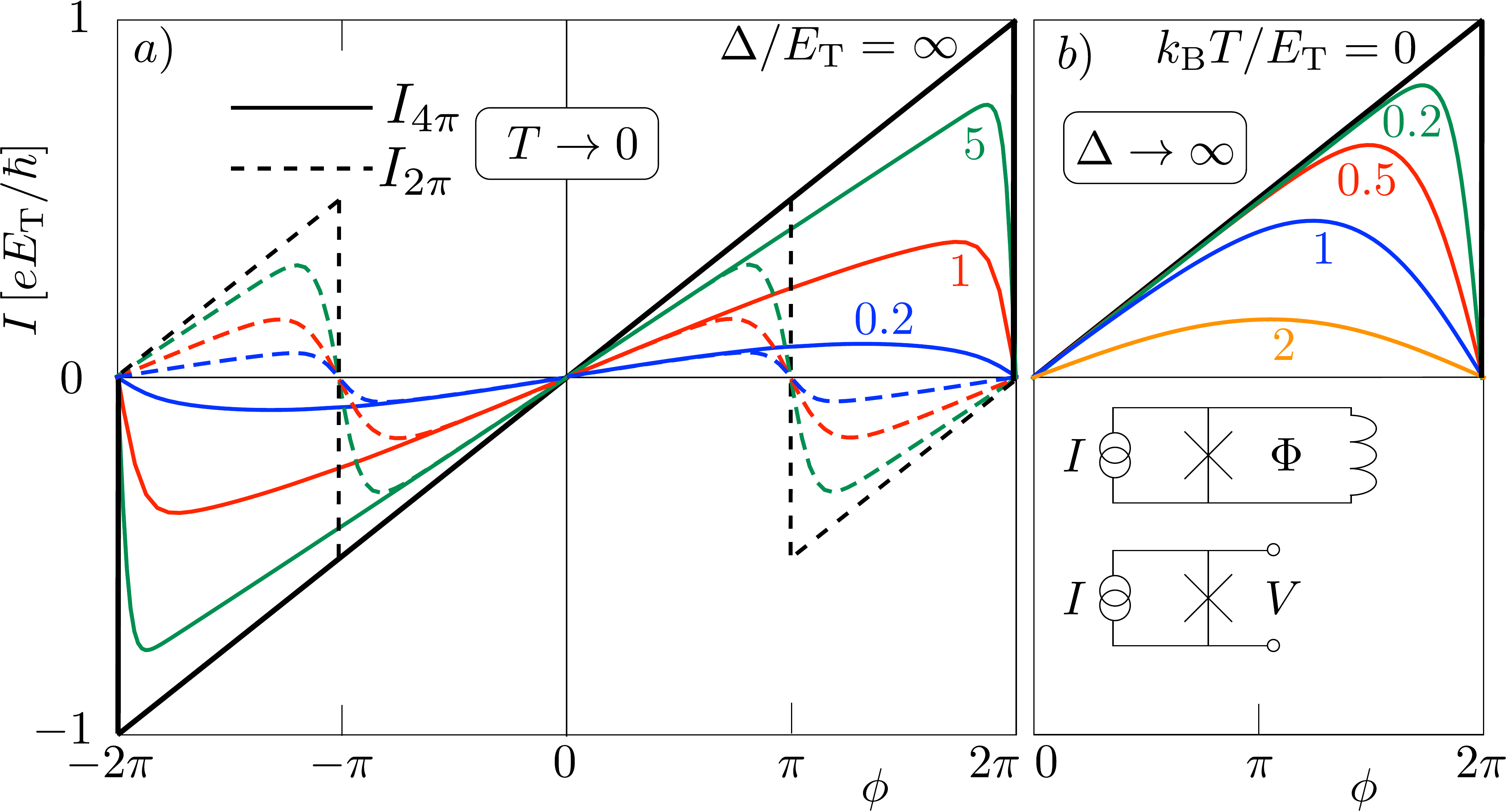}}
\caption{
Phase dependence of the parity-constrained supercurrent $I_{4\pi}$ (solid curves, in units of $eE_{\rm T}/\hbar\propto 1/L$), calculated by a numerical evaluation of the Matsubara sums. The left panel shows the crossover from the short-junction to the long-junction regime in the zero-temperature limit (full interval $-2\pi<\phi<2\pi$). The right panel shows the temperature dependence in the long-junction limit (reduced interval $0<\phi<2\pi$). The left panel also shows the supercurrent $I_{2\pi}$ without parity constraints (dashed curves). The insets in the right panel show current-biased superconducting circuits that measure the $I$-$V$ and $I$-$\Phi$ relationships of a Josephson junction.
}
\label{fig_numerics}
\end{figure}

Putting all pieces together \cite{App} we obtain the parity-dependent supercurrent for arbitrary ratio $\Delta/E_{\rm T}$. In the short-junction limit $\Delta/E_{\rm T}\rightarrow 0$ we recover the known result \eqref{Iphipm}, when the energy dependence of the scattering matrix and the phase sensitivity of the continuous spectrum can both be ignored. In the opposite long-junction limit $\Delta/E_{\rm T}\rightarrow\infty$ we find
\begin{align}
&I_{4\pi}=I_{0}-\frac{2e}{\hbar\beta}\frac{d}{d\phi}\ln\bigl[\tfrac{1}{2}+\cos(\phi/2)e^{{\cal S}-\pi/2\beta E_{\rm T}}\bigr],\label{Ipmresulta}\\
&{\cal S}=\sum_{p=1}^{\infty}(-1)^{p}\ln\bigl(1+2e^{-\Omega_{p}/E_{\rm T}}\cos\phi+e^{-2\Omega_{p}/E_{\rm T}}\bigr),\label{Ipmresultb}\\
&I_{2\pi}\equiv I_{0}=\frac{2e}{\hbar\beta}\sin\phi\sum_{p=0}^{\infty}\bigl[\cos\phi+\cosh(2\omega_{p}/E_{\rm T})\bigr]^{-1}.\label{I0result}
\end{align}
The plot of the results in Fig.\ \ref{fig_numerics} shows that the crossover from a sine to a sawtooth shape occurs early: already for $\Delta=E_{\rm T}$ (so for $L=\xi$) the maximum of the current-phase relationship is close to $\phi=2\pi$. The sawtooth shape is preserved with increasing temperature for $k_{\rm B}T\lesssim\frac{1}{2}E_{\rm T}$.

These are encouraging results for the experimental accessibility of the long-junction regime. The quantum spin-Hall effect has been observed in HgTe/CdTe quantum wells \cite{Kon07}, and more recently in InAs/GaSb quantum wells \cite{Kne11a} --- where also Andreev reflection from superconducting Nb electrodes was demonstrated \cite{Kne11b}. For a typical Fermi velocity of $v\simeq 10^{5}\,{\rm m/s}$ in a semiconductor and superconducting gap $\Delta\simeq 1\,{\rm meV}$ in bulk Nb, the coherence length is $\xi=70\,{\rm nm}$, so the Josephson junction length $L=0.5\,\mu{\rm m}$ from Ref.\ \onlinecite{Kne11b} is deep in the long-junction regime. Since the long-junction regime is already entered for $L\approx\xi$, this would apply even if the effective superconducting gap is well below the bulk value of Nb. The corresponding Thouless energy is $E_{\rm T}/k_{\rm B}=1.5\,{\rm K}$, so at $T=100\,{\rm mK}$ one should be close to the low-temperature limit.

In the ongoing search for the $4\pi$-periodic Josephson effect the first results have been reported \cite{Rok12} for the {\sc ac} effect (fractional Shapiro steps \cite{Kwo03,Bad11,San12,Dom12,Pik11}). A {\sc dc} measurement of the current-flux ($I$-$\Phi$, $\phi=2e\Phi/\hbar$) relationship, on time scales large compared to the time $\tau_{\rm qp}\simeq\mu{\rm s}$ for unpaired quasiparticles to tunnel into the system \cite{Rai12}, will measure the $2\pi$-periodic $I_{2\pi}$ rather than $I_{4\pi}$. Such a phase-sensitive measurement (Fig.\ \ref{fig_numerics}, upper inset) would produce the critical current $I_{2\pi,{\rm c}}$ without any signature of the parity anomaly. In contrast, a phase-\textit{insensitive} measurement of the critical current through the current-voltage ($I$-$V$) characteristic (lower inset) will produce $I_{4\pi,{\rm c}}$ even on time scales $\gg\tau_{\rm qp}$, because the phase of a resistively shunted (overdamped) circuit can adjust to a change in ${\cal N}$ on time scales much smaller than $\tau_{\rm qp}$. A change in the parity of ${\cal N}$ will be compensated by a $2\pi$ phase shift, without a change in critical current \cite{App}. In a short junction $I_{2\pi,{\rm c}}$ and $I_{4\pi,{\rm c}}$ are the same, so this does not help, but in a long junction they differ by up to a factor of two.

In conclusion, we have presented a theory for the $4\pi$-periodic Josephson effect on scales large compared to the superconducting coherence length. A multitude of subgap states, as well as a continuum of states above the gap, contribute to the supercurrent for $L\gg\xi$, but still the parity anomaly responsible for the $4\pi$-periodicity persists. In fact, we have found that in a long junction the anomaly manifests itself also in a phase-insensitive way, through a doubling of the critical current. This opens up new possibilities for the detection of this topological effect at the quantum spin-Hall edge \cite{Kon07,Kne11a,Kne11b}, and possibly also in semiconductor nanowires \cite{Rok12,Mou12,Den12,Das12}.

Discussions with A. R. Akhmerov are gratefully acknowledged. This research was supported by the Dutch Science Foundation NWO/FOM and by an ERC Advanced Investigator grant.

\appendix

\section{Details of the calculation of the free energy}
\label{AppA}

\subsection{Transformation from the real to the imaginary energy axis}
\label{conversion}

According to Eq.\ \eqref{Zpmdef} the free energy $F_{\sigma}=-\beta^{-1}\ln Z_{\sigma}$, with $\sigma=\pm 1$ the ground-state fermion parity, is given by
\begin{align}
F_{\sigma}&=F_{0}-\beta^{-1}\ln\tfrac{1}{2}\biggl[1+\sigma\exp\biggl(\sum_{\varepsilon>0}\ln\tanh(\beta\varepsilon/2)\biggr)\biggr],\label{FpmdefappA}\\
F_{0}&=-\beta^{-1}\sum_{\varepsilon>0}\ln [2\cosh(\beta\varepsilon/2)].\label{F0defapA}
\end{align}
Here $F_{0}$ is the free energy in the absence of parity constraints. The infinite product over $\varepsilon$ is defined in terms of the density of states $\rho$ by
\begin{equation}
\prod_{\varepsilon>0}f(\varepsilon)=\exp\biggl[\int_{0}^{\infty}d\varepsilon\,\rho(\varepsilon)\ln f(\varepsilon)\biggr].\label{prodsumrelation}
\end{equation}

Integrals of the type \eqref{prodsumrelation} can be done efficiently by contour integration, made possible by the fact that the scattering matrices are analytic in the upper half of the complex energy plane. This was worked out in Ref.\ \onlinecite{Bro97} for $F_{0}$, and here we adapt that method to include the parity constraint in $F_{\sigma}$.

The function $\nu$ in the density of states \eqref{rhorhoSdef} satisfies $\nu(\varepsilon)=\nu^{\ast}(-\varepsilon)$, expressing the electron-hole symmetry. If $f(\varepsilon)$ is an even function of $\varepsilon$, we may therefore convert the sum $\sum_{\varepsilon>0}f(\varepsilon)$ over positive energies (including both the discrete and the continuous spectrum) into an integral along the entire real energy axis of $fd\nu/d\varepsilon$, or $\nu df/d\varepsilon$ after a partial integration. Closing the contour in the upper half of the complex energy plane picks up the poles of $df/d\varepsilon$, which for the free energy are the Matsubara frequencies on the imaginary axis.

In Ref.\ \onlinecite{Bro97} this transformation was carried out for $F_{0}$, leading to
\begin{align}
&\sum_{\varepsilon>0}\ln [2\cosh(\beta\varepsilon/2)]=\int_{0}^{\infty}d\varepsilon\,\rho(\varepsilon)\ln [2\cosh(\beta\varepsilon/2)]\nonumber\\
&\quad\quad\quad=\frac{1}{2i}\int_{-\infty}^{\infty}d\varepsilon\,\frac{d\nu(\varepsilon+i0^{+})}{d\varepsilon}\ln [2\cosh(\beta\varepsilon/2)]\nonumber\\
&\quad\quad\quad=-\frac{\beta}{4i}\int_{-\infty}^{\infty}d\varepsilon\,\nu(\varepsilon+i0^{+})\tanh(\beta\varepsilon/2)\nonumber\\
&\quad\quad\quad=-\pi\sum_{p=0}^{\infty}\nu(i\omega_{p}),\label{coshsum}
\end{align}
with $\omega_{p}=(2p+1)\pi/\beta$ the fermionic Matsubara frequency. For $F_{\sigma}$ the Matsubara sum contains bosonic Matsubara frequencies $\Omega_{p}=2p\pi/\beta$,
\begin{align}
&\sum_{\varepsilon>0}\ln\tanh(\beta\varepsilon/2)=\frac{1}{2i}\int_{-\infty}^{\infty}d\varepsilon\,\frac{d\nu(\varepsilon+i0^{+})}{d\varepsilon}\nonumber\\
&\quad\quad\quad\quad\quad\quad\quad\quad\quad\times\ln|\tanh(\beta\varepsilon/2)|\nonumber\\
&\quad\quad\quad=-\frac{\beta}{2i}\fint_{-\infty}^{\infty}d\varepsilon\,\nu(\varepsilon+i0^{+})\frac{1}{\sinh\beta\varepsilon}\nonumber\\
&\quad\quad\quad=-\pi\sum_{p=1}^{\infty}(-1)^{p}\nu(i\Omega_{p}/2)-\tfrac{1}{2}\pi\nu(0).\label{sinhsum}
\end{align}
The notation $\fint$ indicates the Cauchy principal value of the integral (which picks up half of the pole at $\varepsilon=0$). We have made use of the identity
\begin{equation}
\int dx\,\frac{df}{dx}\ln|x|=-\fint dx\,f(x)\frac{1}{x}.
\end{equation}

For the free energy this gives
\begin{widetext}
\begin{align}
F_{0}&=-\beta^{-1}\sum_{p=0}^{\infty}\ln{\rm Det}\,X(i\omega_{p}),\label{F0def}\\
F_{\sigma}&=F_{0}-\beta^{-1}\ln\tfrac{1}{2}\left[1+\sigma e^{J_{\rm S}}\sqrt{{\rm Det}\,X(0)}\exp\left(\sum_{p=1}^{\infty}(-1)^{p}\ln{\rm Det}\,X(i\Omega_{p}/2)\right)\right],\label{Fpmdef}
\end{align}
\end{widetext}
where we have substituted $\nu=-\pi^{-1}\ln{\rm Det}\,X$ from Eq.\ \eqref{nudef} and also included the factor $e^{J_{\rm S}}$ from the $\phi$-independent density of states in the superconducting electrodes.

\subsection{Regularization}
\label{regularize}

The transformation of the integral \eqref{sinhsum} over real energies into a sum over imaginary frequencies requires that $\nu(i\omega)$ goes to zero faster than $1/\omega$ for $\omega\rightarrow\infty$. To ensure this, we decompose $\nu=\nu_{\infty}+\delta\nu$, with $\nu_{\infty}$ the large-$\omega$ limit of $\nu(i\omega)$. It is convenient to specify $\nu_{\infty}(0)=0$. The integral over $\nu_{\infty}$ is done along the real energy axis, where it converges, and then the remaining integral over $\delta\nu$ becomes a converging sum over Matsubara frequencies.

More specifically, for the quantum spin-Hall edge we take
\begin{equation}
\nu_{\infty}(\varepsilon)=-\pi^{-1}\ln\bigl[(1-4\varepsilon^2/\Delta^{2})e^{-2i\varepsilon/E_{\rm T}}\bigr],\label{nuinftydef}
\end{equation}
in view of Eq.\ \eqref{DetXresult}. The integral over $\nu_{\infty}$ can be evaluated in closed form,
\begin{align}
J_{\infty}&\equiv-\frac{\beta}{2i}\fint_{-\infty}^{\infty}d\varepsilon\,\nu_{\infty}(\varepsilon+i0^{+})\frac{1}{\sinh\beta\varepsilon}\nonumber\\
&=-\frac{1}{\pi \beta E_{\rm T}}\int_{-\infty}^{\infty}dx\,\frac{x}{\sinh x}-\int_{\beta\Delta/2}^{\infty}dx\,\frac{1}{\sinh x}\nonumber\\
&=-\frac{\pi}{2\beta E_{\rm T}}+\ln\tanh(\beta\Delta/4).\label{nuinftyint}
\end{align}

The remaining integral over $\delta\nu=\nu-\nu_{\infty}$ then becomes a convergent sum over Matsubara frequencies,
\begin{align}
&-\frac{\beta}{2i}\fint_{-\infty}^{\infty}d\varepsilon\,\frac{\delta\nu(\varepsilon+i0^{+} )}{\sinh\beta\varepsilon}={\cal S}-\tfrac{1}{2}\pi\nu(0),\\
&{\cal S}=-\pi\sum_{p=1}^{\infty}(-1)^{p}\bigl[\nu(i\Omega_{p}/2)-\nu_{\infty}(i\Omega_{p}/2)\bigr].\label{calSdef}
\end{align}
This gives the regularized version of Eq.\ \eqref{Fpmdef},
\begin{equation}
F_{\sigma}=F_{0}-\beta^{-1}\ln\tfrac{1}{2}\bigl(1+\sigma e^{{\cal S}+J_{\infty}+J_{\rm S}}\sqrt{{\rm Det}\,X(0)}\,\bigr).\label{FF0formulareg}
\end{equation}

\section{Scattering formulas for the ground-state fermion parity}
\label{scatteringsigma}

The ground-state fermion parity $\sigma(\phi)$ switches between even and odd whenever a pair of Andreev levels crosses the Fermi energy. Given $\sigma(0)=1$, in principle one can just count the number of level crossings between phase difference $0$ and $\phi$ to determine $\sigma$. In a multichannel Josephson junction there can be many level crossings and a single crossing might be easily missed. It would be advantageous to have a direct method of determining $\sigma$ at any $\phi$, without having to track the number of sign changes back to $\phi=0$. Kitaev's Hamiltonian expression \cite{Kit01} for the ground-state fermion parity is one such method, requiring information on the entire excitation spectrum. Here we construct an alternative scattering approach that requires only Fermi-level information.

We present two variations of our approach: the first assumes spatial separation of normal scattering and Andreev reflection, relating $\sigma$ to the normal-state scattering matrix $s_{0}$. Alternatively, if there is no such spatial separation, we can relate $\sigma$ to the transfer matrices $M_{\rm L}$, $M_{\rm R}$ at the left and right end of the Josephson junction.

It is instructive to place these results in the general context of topological states of matter \cite{Ryu10}. The ground-state fermion parity $\sigma$ is the $\mathbb{Z}_{2}$ topological quantum number of a system of dimensionality $d=0$ in symmetry class D (when $s_{0}$ has no symmetry restrictions) or BDI (when $s_{0}=s_{0}^{\rm T}$). The dimensionality zero refers to the fact that this is a closed system. We may consider opening up the system, promoting it to $d=1$, by replacing one of the two superconducting contacts by a normal metal with $N$ transverse modes. The topological quantum number ${\cal Q}$ then counts the number of Majorana zero-modes at the normal-superconducting interface. It is given by the determinant of the reflection matrix (for class D, with ${\cal Q}\in\mathbb{Z}_{2}$) \cite{Akh11} or by its trace (for class BDI, with ${\cal Q}\in\mathbb{Z}$) \cite{Die12}.

\subsection{Relation between $\bm{\sigma}$ and the normal-state scattering matrix}
\label{sigmas0rA}

The free energy \eqref{Fpmdef} of the SNS junction incorporates the ground-state fermion parity dependence through the quantity $\sigma\sqrt{{\rm Det}\,X(0)}$. We seek to express this quantity in terms of the normal-state scattering matrix $s_{0}(0)$, assuming a spatial separation of normal scattering in N and ideal Andreev reflection at the NS interfaces.

We start from the definition of $X=(1-M)M^{-1/2}$ and use that $M(0)=U^{\ast}U$ with unitary $U=r_{\rm A}(0)s_{0}(0)$. Since ${\rm Det}\,M=1$, we have
\begin{align}
{\rm Det}\,X&={\rm Det}\,(1-U^{\ast}U)={\rm Det}\,(1-U^{\ast}U)^{\rm T}\nonumber\\
&=\frac{{\rm Det}\,(U-U^{\rm T})}{{\rm Det}\,U}=\frac{[{\rm Pf}(U-U^{\rm T})]^{2}}{{\rm Det}\,U}.\label{DetXPfU}
\end{align}
The determinant of $U$ is independent of $\phi$,
\begin{equation}
{\rm Det}\,U={\rm Det}\,[r_{\rm A}(0)s_{0}(0)]={\rm Det}\,[is_{0}(0)].\label{DetU}
\end{equation}

We may now identify
\begin{align}
&\sigma\sqrt{{\rm Det}\,X(0)}=\frac{{\rm Pf}\,(U^{\rm T}-U)}{\sqrt{{\rm Det}\,U}}\nonumber\\
&\Rightarrow\sigma={\rm sign}\,\left[\frac{{\rm Pf}\,(r_{\rm A}s_{0}-s_{0}^{\rm T}r_{\rm A}^{\rm T})}{\sqrt{{\rm Det}\,is_{0}}}\right]_{\varepsilon=0}.\label{sigmaDetX}
\end{align}
The two branches of the square root function introduce a sign ambiguity, which is resolved as follows. The square root of ${\rm Det}\,X\geq 0$ is taken on the principal branch, while the branch of the square root of ${\rm Det}\,is_{0}$ is fixed by setting $\sigma=1$ at $\phi=0$. Once this sign is fixed, the phase dependence of the topological quantum number $\sigma(\phi)$ is determined by Eq.\ \eqref{sigmaDetX} entirely in terms of Fermi-level properties.

\subsection{Relation between $\bm{\sigma}$ and the transfer matrix}
\label{sigmaMLMR}

We will now extend the scattering formulation of the ground-state fermion parity to a system where we cannot make the spatial separation of normal scattering (described by $s_0$) and ideal Andreev reflection (described by $r_{\rm A}$). This is possible if we work with transfer matrices instead of scattering matrices.

It is convenient to choose a gauge where the phase difference $\phi$ across the Josephson junction is accounted for by the delta function vector potential $\vec{A}=(\phi\hbar/2e)\delta(x)\hat{x}$, centered at a point $x=0$ inside the Josephson junction. The $2N\times 2N$ transfer matrices $M_{\rm L}(\varepsilon)$ and $M_{\rm R}(\varepsilon)$ relate electron and hole wave amplitudes to the left of $x=0$,
\begin{equation}
\Psi_{\rm L,h} = M_{\rm L} \Psi_{\rm L,e},\label{MLdef}
\end{equation}
and to the right of $x=0$,
\begin{equation}
\Psi_{\rm R,h} = M_{\rm R} \Psi_{\rm R,e}.\label{MRdef}
\end{equation}
The first $N$ components $\Psi_{+}$ of each vector $\Psi=(\Psi_{+},\Psi_{-})$ refer to right-moving states and the last $N$ components $\Psi_{-}$ to left-moving states. The Pauli matrix $\Sigma_{z}$ acts on these components,
\begin{equation}
\Sigma_{z}\begin{pmatrix}
\Psi_{+}\\
\Psi_{-}
\end{pmatrix}=
\begin{pmatrix}
1&0\\
0&-1
\end{pmatrix}\begin{pmatrix}
\Psi_{+}\\
\Psi_{-}
\end{pmatrix}=
\begin{pmatrix}
\Psi_{+}\\
-\Psi_{-}
\end{pmatrix}.\label{Sigmazdef}
\end{equation}

The wave amplitudes $\Psi_{\rm L}$ and $\Psi_{\rm R}$ are matched at $x=0$,
\begin{equation}
\Psi_{\rm R,e} = e^{i\phi/2}\Psi_{\rm L,e},\;\;
\Psi_{\rm R,h} = e^{-i\phi/2}\Psi_{\rm L,h}.\label{Psimatching}
\end{equation}
The combination of these equations gives
\begin{equation}
e^{i\phi/2} M_{\rm R}\Psi_{\rm L,e} = e^{-i\phi/2} M_{\rm L} \Psi_{\rm L,e},\label{MRML}
\end{equation}
so the condition for a bound state in the junction is
\begin{equation}
{\rm Det}\left[ e^{i\phi/2}M_{\rm R} - e^{-i\phi/2} M_{\rm L}\right]=0.\label{MRMLdet}
\end{equation}
If we take $M_{\rm L}$ and $M_{\rm R}$ at the Fermi energy $\varepsilon=0$, this equation gives the values of $\phi$ at which a pair of Andreev levels crosses the Fermi level and the ground-state fermion parity switches between even and odd.

Because of the excitation gap in the bulk superconductor, there can be no particle current flowing through the Josephson junction for energies $\varepsilon<\Delta$. This requires
\begin{equation}
\Psi_{\rm e}^{\dagger}\Sigma_{z}\Psi_{\rm e}+\Psi_{\rm h}^{\dagger}\Sigma_{z}\Psi_{\rm h}=0,\label{PsiePsih}
\end{equation}
both to the left and to the right of $x=0$. The corresponding unitarity constraint on the transfer matrices $M_{\rm L}$ and $M_{\rm R}$ is
\begin{equation}
M^{\dagger}\Sigma_{z} M+\Sigma_{z}=0\Rightarrow M^{-1}=-\Sigma_{z} M^{\dagger}\Sigma_{z}.\label{Munitary}
\end{equation}
At the Fermi level $\varepsilon=0$ we have the additional constraint of particle-hole symmetry, $M^{-1}(0)=M^{\ast}(0)$, which together with the unitarity constraint implies that $M(0)\Sigma_{z}$ is an antisymmetric matrix,
\begin{equation}
[M(0)\Sigma_{z}]^{\rm T}=-M(0)\Sigma_{z}.\label{Mehsymm}
\end{equation}
In what follows we restrict ourselves to $\varepsilon=0$ and omit the energy argument.

Since $|{\rm Det}\,M|=1$, we may define the real angle $\alpha$ (modulo $2\pi$) by
\begin{equation}
{\rm Det}\,(M_{\rm L}M_{\rm R})=e^{-i\alpha-2i\nu\pi},\;\;0\leq \alpha<2\pi,\;\;\nu\in\mathbb{Z}.\label{DetMLMR}
\end{equation}
The function
\begin{equation}
\zeta^{2}(\phi)=e^{i\alpha/2+i\nu\pi}\,{\rm Det}\,\left( e^{i\phi/2}M_{\rm R}\Sigma_{z} - e^{-i\phi/2} M_{\rm L}\Sigma_{z}\right)\label{zeta2}
\end{equation}
is \textit{real} for all $\phi$ and vanishes when the ground-state fermion parity switches. Since it is the determinant of an antisymmetric matrix, it can be written as the square of a Pfaffian,
\begin{equation}
\zeta(\phi)=e^{i\alpha/4+i\nu\pi/2}\,{\rm Pf}\,\left( e^{i\phi/2}M_{\rm R}\Sigma_{z} - e^{-i\phi/2} M_{\rm L}\Sigma_{z}\right).\label{zetaphi}
\end{equation}

We choose $\nu\in\{0,1,2,3\}$ such that $\zeta(0)$ is real and positive. The function $\zeta(\phi)$ then will remain real for all $\phi$, switching sign when the ground-state fermion parity switches. We can thus identify the topological quantum number with
\begin{equation}
\sigma={\rm sign}\,\left[e^{i\alpha/4+i\nu\pi/2}{\rm Pf}\left( e^{i\phi/2}M_{\rm R}\Sigma_{z} - e^{-i\phi/2} M_{\rm L}\Sigma_{z}\right)\right].\label{sigmaphisign}
\end{equation}

\subsection{Multichannel applications}
\label{multiNQ}

In the main text we apply our scattering formulation to the quantum spin-Hall edge, which has $N=1$ channels (counting spin) and ${\cal Q}=1$ Majorana zero-modes (at each NS interface). More generally, the formulas given apply directly to any $N\geq 1$, with the requirement that $N-{\cal Q}$ is an even integer. This is a technical requirement, to avoid the difficulty that for $N-{\cal Q}$ odd one of the $N$ channels has identically zero Andreev reflection probability at the Fermi level \cite{Die12}. Here we show how this restriction can be removed, first in terms of the scattering matrix, then in terms of the transfer matrix.

\subsubsection{Scattering matrix formulation}

We assume a spatial separation of normal-state scattering (described by $s_{0}$) and Andreev reflection (described by $r_{\rm A}$). For $N-{\cal Q}$ odd one channel is fully decoupled from the NS interface, so we cannot include it in $r_{\rm A}$. We are free to choose basis states such that the decoupled channel has the index $N$ (for the left interface) and $2N$ (for the right interface). Our goal is to determine the reduced unitary scattering matrix $\tilde{s}_0$ that relates the incoming and outgoing electrons in the remaining channels. This is obtained from the relation $\psi^{\rm out}=s_0\psi^{\rm in}$ by algebraic elimination of the decoupled components $\psi^{\rm out}_N=\psi^{\rm in}_N$ and $\psi^{\rm out}_{2N}=\psi^{\rm in}_{2N}$. We thus arrive at
\begin{equation}
\tilde{s}_{0}=P^{\rm T}s_{0}[1-(1-PP^{\rm T})s_{0}]^{-1}P,\label{tildes0}
\end{equation}
with the $2N\times (2N-2)$-dimensional matrix
\begin{equation}
P_{nm}=\begin{cases}
\delta_{n,m}&{\rm if}\;\;1\leq n \leq N-1\\
\delta_{n,m+1}&{\rm if}\;\;N+1\leq n \leq 2N-1\\
0&{\rm if}\;\;n\in\{N,2N\} 
\end{cases}.
\end{equation}
The combination $1-PP^{\rm T}$ projects onto the decoupled channels. The remaining channels are Andreev reflected with unit probability, as described by the  $2(N-1)\times 2(N-1)$-dimensional unitary matrix $r_{\rm A}$. All formulas carry through, with the replacement of $s_{0}$ by $\tilde{s}_{0}$.

With appropriately chosen $P$, this construction extends to cases with multiple decoupled channels, including situations where their number differs at the two interfaces.

\subsubsection{Transfer matrix formulation}

We again choose a particular set of basis states for incoming and outgoing modes at the left and right interface, such that the decoupled channel has index $N$ and $2N$, respectively. An electron in this channel is reflected as an electron and a hole is reflected as a hole. Therefore these channels do not appear in the $2(N-1)\times 2(N-1)$ electron-to-hole transfer matrices $\tilde{M}_{\rm L}$ and $\tilde{M}_{\rm R}$. These relate
\begin{equation}
\tilde{\Psi}_{\rm L,h} = \tilde{M}_{\rm L} \tilde{\Psi}_{\rm L,e},\;\;
\tilde{\Psi}_{\rm R,h} = \tilde{M}_{\rm R} \tilde{\Psi}_{\rm R,e},\label{tildeMrelation}
\end{equation}
where $\tilde{\Psi}$ differs from $\Psi$ because it does not include the decoupled channel.  

When we match wave functions at $x=0$ we have to take into account that the basis states need not coincide: the basis that decouples the channel from the left interface can be different from the basis that decouples it from the right interface. The matching condition \eqref{Psimatching} thus contains a pair of $N\times N$ unitary matrices $u_{\rm match},v_{\rm match}$ to change the basis of left-moving and right-moving states,
\begin{equation}
\begin{split}
&\Psi_{\rm R,e} = e^{i\phi/2}\begin{pmatrix}
u_{\rm match}&0\\
0&v_{\rm match}
\end{pmatrix}\Psi_{\rm L,e},\\
&\Psi_{\rm R,h} = e^{-i\phi/2}\begin{pmatrix}
u^{\ast}_{\rm match}&0\\
0&v^{\ast}_{\rm match}
\end{pmatrix}\Psi_{\rm L,h}.
\end{split}
\label{Psimatchingnew}
\end{equation}

The matching matrices $u_{\rm match},v_{\rm match}$ correspond to a $2N\times 2N$ unitary scattering matrix $s_{\rm match}$ that has only transmission blocks,
\begin{equation}
s_{\rm match}=\begin{pmatrix}
0&u_{\rm match}\\
v_{\rm match}^{\dagger}&0
\end{pmatrix}.
\end{equation}
We apply the projection \eqref{tildes0} to $s_{\rm match}$ to obtain a $2(N-1)\times 2(N-1)$ unitary scattering matrix $\tilde{s}_{\rm match}$ that does not contain the decoupled channels. We then transform back from scattering matrix $\tilde{s}_{\rm match}$ to transfer matrix $\tilde{m}_{\rm match}$, which relates
\begin{equation}
\tilde{\Psi}_{\rm R,e}=e^{i\phi/2}\tilde{m}_{\rm match}\Psi_{\rm L,e},\;\;
\tilde{\Psi}_{\rm R,h}=e^{-i\phi/2}\tilde{m}_{\rm match}^{\ast}\Psi_{\rm L,h}.\label{PsitildeMmatch}
\end{equation}
Unitarity of $\tilde{s}_{\rm match}$ implies for $\tilde{m}_{\rm match}$ the condition
\begin{equation}
\tilde{m}_{\rm match}^{-1}=\Sigma_{z} \tilde{m}_{\rm match}^{\dagger}\Sigma_{z}.\label{mmatchunitary}
\end{equation}

Combining Eqs.\ \eqref{tildeMrelation} and \eqref{PsitildeMmatch} we arrive at
\begin{equation}
e^{i\phi/2} \tilde{M}_{\rm R}\tilde{m}_{\rm match}\tilde{\Psi}_{\rm L,e} = e^{-i\phi/2} \tilde{m}_{\rm match}^{\ast}\tilde{M}_{\rm L} \tilde{\Psi}_{\rm L,e},\label{MRMLnew}
\end{equation}
and the corresponding condition for a bound state,
\begin{equation}
{\rm Det}\left[ e^{i\phi/2}\tilde{M}_{\rm R}\tilde{m}_{\rm match} - e^{-i\phi/2} \tilde{m}_{\rm match}^{\ast}\tilde{M}_{\rm L}\right]=0.\label{MRMLdetnew}
\end{equation}

We rewrite this using Eq.\ \eqref{mmatchunitary} as the determinant of an \textit{antisymmetric} matrix,
\begin{equation}
{\rm Det}\left[ e^{i\phi/2}\tilde{M}_{\rm R}\Sigma_{z} - e^{-i\phi/2}\tilde{m}_{\rm match}^{\ast}\tilde{M}_{\rm L}\Sigma_{z}\tilde{m}_{\rm match}^{\dagger}\right]=0.\label{MRMLdetnewer}
\end{equation}
We then proceed as in Eqs.\ \eqref{DetMLMR}--\eqref{sigmaphisign}, with $M_{\rm R}$ replaced by $\tilde{M}_{\rm R}$ and $M_{\rm L}$ replaced by $\tilde{m}_{\rm match}^{\ast}\tilde{M}_{\rm L}\Sigma_{z}\tilde{m}_{\rm match}^{\dagger}\Sigma_{z}$.

\section{Evaluation of the supercurrent along the quantum spin-Hall edge}
\label{evaluation}

We apply Eq.\ \eqref{FF0formulareg} to the quantum spin-Hall edge. According to Eq.\ \eqref{DetXresult}, we have
\begin{align}
&\nu(i\omega)=-\frac{1}{\pi}\ln\bigl[2\cos\phi+\zeta_{+}(\omega)+\zeta_{-}(\omega)\bigr],\label{nuomegaQSH}\\
&\zeta_{\pm}(\omega)=e^{\pm 2\omega/E_{\rm T}}\bigl(\sqrt{1+\omega^{2}/\Delta^{2}}\pm\omega/\Delta\bigr)^{2}.\label{zetapmdef}
\end{align}
The Matsubara sum \eqref{calSdef}, with $\nu_{\infty}$ given by Eq.\ \eqref{nuinftydef}, takes the form
\begin{equation}
{\cal S}=\sum_{p=1}^{\infty}(-1)^{p}\ln\biggl[\frac{2\cos\phi+\zeta_{+}(\Omega_{p}/2)+\zeta_{-}(\Omega_{p}/2)}{(1+\Omega_{p}^2/\Delta^{2})e^{\Omega_{p}/E_{\rm T}}}\biggr]\label{calSQSH}
\end{equation}
The function $J_{\infty}$ is given by Eq.\ \eqref{nuinftyint} and
\begin{equation}
\sigma\sqrt{{\rm Det}\,X(0)}=2\cos(\phi/2),
\end{equation}
in view of Eqs.\ \eqref{s0QSH} and \eqref{sigmaDetX}.

The superconducting electrodes couple to the quantum spin-Hall edge via a single transverse mode, over a total length $L_{\rm S}$. The corresponding density of states is
\begin{equation}
\rho_{\rm S}(\varepsilon)=\frac{2}{\pi E_{\rm S}}\frac{|\varepsilon|}{\sqrt{\varepsilon^{2}-\Delta^{2}}},\;\;|\varepsilon|>\Delta.
\end{equation}
We have defined $E_{\rm S}=\hbar v_{\rm F}/L_{\rm S}$. The superconducting electrodes affect the parity-dependent free energy \eqref{FF0formulareg} through the factor $e^{J_{\rm S}}$, with
\begin{align}
J_{\rm S}&=\int_{\Delta}^{\infty}d\varepsilon\,\rho_{\rm S}(\varepsilon)\ln\tanh(\beta\varepsilon/2)\nonumber\\
&=-\frac{2\beta}{\pi E_{\rm S}}\int_{\Delta}^{\infty}d\varepsilon\,\frac{\sqrt{\varepsilon^{2}-\Delta^{2}}}{\sinh\beta\varepsilon}.
\end{align}

Collecting results, we arrive at the parity-dependent supercurrent
\begin{align}
&I_{\sigma}=I_{0}-\frac{2e}{\hbar\beta}\frac{d}{d\phi}\ln\tfrac{1}{2}\bigl[1+\sigma|\cos(\phi/2)|e^{{\cal S}+J_{\infty}+J_{\rm S}}\bigr],\label{Ipmgeneral}\\
&\sigma={\rm sign}\,[\cos(\phi/2)],\label{sigmaQSH}
\end{align}
with $I_{0}$ the $2\pi$-periodic supercurrent in the absence of parity constraints,
\begin{equation}
I_{0}=\frac{4e}{\hbar\beta}\sin\phi\sum_{p=0}^{\infty}[2\cos\phi+\zeta_{+}(\omega_{p})+\zeta_{-}(\omega_{p})]^{-1}.\label{I0general}
\end{equation}
In the long-junction regime $\Delta\gg E_{\rm T},\, k_{\rm B}T$ these results reduce to the equations \eqref{Ipmresulta}--\eqref{I0result} given in the main text.

\subsection{Short-junction limit}
\label{shortjunctionlimit}

As a check on the consistency of the whole formalism, we take the short-junction limit $E_{\rm T}\rightarrow\infty$ of the parity-dependent supercurrent \eqref{Ipmgeneral} and see if we recover the results of Fu and Kane \cite{Fu09}. We choose the phase interval $|\phi|<\pi$ and abbreviate
\begin{equation}
u\equiv\tfrac{1}{2}\cos(\phi/2)\beta\Delta.\label{udef}
\end{equation}

The Matsubara sums \eqref{calSQSH} and \eqref{I0general} can be evaluated in closed form in the short-junction limit,
\begin{align}
\lim_{\rm E_{\rm T}\rightarrow\infty}{\cal S}&=\sum_{p=1}^{\infty}(-1)^{p}\ln\biggl(\frac{2\cos\phi+2+\Omega_{p}^{2}/\Delta^{2}}{1+\Omega_{p}^2/\Delta^{2}}\biggr)\nonumber\\
&=\ln\biggl(\frac{\tanh u}{2\cos(\phi/2)\tanh(\beta\Delta/4)}\biggr),\label{calSshort}\\
\lim_{\rm E_{\rm T}\rightarrow\infty}I_{0}&=\frac{4e}{\hbar\beta}\sin\phi\sum_{p=0}^{\infty}[2\cos\phi+2+4\omega_{p}^{2}/\Delta^{2}]^{-1}\nonumber\\
&=\frac{e\Delta}{2\hbar}\sin(\phi/2)\tanh u.\label{I0short}
\end{align}
In the same limit $J_{\infty}=\ln\tanh(\beta\Delta/4)$; upon substitution into Eq.\ \eqref{Ipmgeneral} we arrive at
\begin{align}
I_{\pm}&=I_{0}-\frac{2e}{\hbar\beta}\frac{d}{d\phi}\ln\bigl(\tfrac{1}{2}\pm\tfrac{1}{2}e^{J_{\rm S}}\tanh u\bigr)\nonumber\\
&=-\frac{2e}{\hbar\beta}\frac{d}{d\phi}\ln\bigl(\cosh u\pm e^{J_{\rm S}}\sinh u\bigr).\label{Ipmshort}
\end{align}

In the zero-temperature limit $J_{\rm S}\rightarrow 0$ and we recover the result of Ref.\ \onlinecite{Fu09},
\begin{equation}
\lim_{T\rightarrow 0}I_{\pm}=\mp\frac{2e}{\hbar\beta}\frac{du}{d\phi}=\pm\frac{e\Delta}{2\hbar}\sin(\phi/2).\label{IzeroTresult}
\end{equation}
 The parity dependence at finite temperature can be quantified by the difference $\delta I=\tfrac{1}{2}(I_{+}-I_{-})$, for which we find the compact expression
\begin{equation}
\delta I=-\frac{4e}{\hbar\beta}\frac{\tau}{1-\tau^{2}}\frac{1}{\sinh 2u}\frac{du}{d\phi},\;\;\tau=e^{J_{\rm S}}\tanh u,\label{deltaIresult}
\end{equation}
in agreement with Ref.\ \onlinecite{Ios11}.

\subsection{Zero-temperature limit in the long-junction regime}
\label{longjunctionlimit}

Another check on the formalism is provided by the combined zero-temperature and long-junction limits. We again choose the interval $|\phi|<\pi$. The Matsubara sums \eqref{calSQSH} and \eqref{I0general} are given in the long-junction limit by
\begin{align}
&\lim_{\Delta\rightarrow\infty}{\cal S}=\sum_{p=1}^{\infty}(-1)^{p}\ln\bigl(1+2e^{-\Omega_{p}/E_{\rm T}}\cos\phi+e^{-2\Omega_{p}/E_{\rm T}}\bigr),\label{calSlong}\\
&\lim_{\Delta\rightarrow\infty}I_{0}=\frac{2e}{\hbar\beta}\sin\phi\sum_{p=0}^{\infty}\bigl[\cos\phi+\cosh(2\omega_{p}/E_{\rm T})\bigr]^{-1}.\label{I0long}
\end{align}

In the zero-temperature limit the sums may be converted into integrals, with the results
\begin{equation}
{\cal S}\rightarrow-\ln|2\cos(\phi/2)|,\;\;
I_{0}\rightarrow\frac{eE_{\rm T}}{2\pi\hbar}\,\phi.\label{limTDelta}
\end{equation}
The two terms $J_{\infty}$ and $J_{\rm S}$ both vanish at $T=0$. Substitution into Eq.\ \eqref{Ipmgeneral} gives $I_{+}=I_{0}$, in agreement with Eq.\ \eqref{Iphipmlong}, while $I_{-}$ remains undetermined. The zero-temperature limit of $I_{-}$ depends on on higher order terms in the low-temperature expansion of ${\cal S}$ that we have not been able to calculate analytically. A numerical calculation (using the Pad\'{e} approximant built into the {\tt NSum|AlternatingSigns} routine of {\sc Mathematica}) gives
\begin{equation}
\lim_{T\rightarrow 0}\lim_{\Delta\rightarrow\infty}\frac{2}{\beta E_{\rm T}}\ln\biggl(\tfrac{1}{2}-|\cos(\phi/2)|e^{{\cal S}+J_{\infty}}\biggr)=|\phi|-\pi,\label{limTzero}
\end{equation}
resulting in a current $I_{-}$ in agreement with Eq.\ \eqref{Iphipmlong}.

\section{Circuits to measure the critical current with and without parity constraints}
\label{AppB}

As explained in the main text, the two circuits shown in Fig.\ \ref{fig_numerics} (inset) both measure the critical current of the Josephson junction, but their sensitivity to fermion-parity constraints is fundamentally different. A measurement of the current-phase relationship (upper circuit) is insensitive to parity constraints when quasiparticles can enter or leave the system on the time scale of the measurement. This gives the critical current $I_{2\pi,{\rm c}}$. A measurement of the current-voltage characteristic (lower circuit) remains governed by fermion-parity constraints as long as the quasiparticle tunneling time $\tau_{\rm qp}$ is large compared to the phase relaxation time $\tau_{\rm J}$ of the resistively shunted Josephson junction. This then gives $I_{4\pi,{\rm c}}$. Here we analyze these two circuits in some more detail.

In the zero-temperature, long-junction limit, we have the $4\pi$-periodic sawtooth current-phase relationship shown in Fig.\ \ref{fig_spectrum} (lower panel). This plot is for an even number of electrons in the system, ${\cal P}\equiv (-1)^{\cal N}=1$, while for odd parity ${\cal P}=-1$ the sawtooth is displaced horizontally by $2\pi$. Both cases are contained in the formula
\begin{equation}
I_{\cal P}(\phi)=\frac{I_{4\pi,{\rm c}}}{2\pi}{\rm mod}_{4\pi}(\phi+{\cal P}\pi+\pi)-I_{4\pi,{\rm c}}.\label{current-phase}
\end{equation}
The modulo function is defined by ${\rm mod}_{4\pi}(\phi)=\phi-4\pi n$, with $n\in\mathbb{Z}$ such that ${\rm mod}_{4\pi}(\phi)\in[0,4\pi)$.

\begin{figure}[tb]
\centerline{\includegraphics[width=0.8\linewidth]{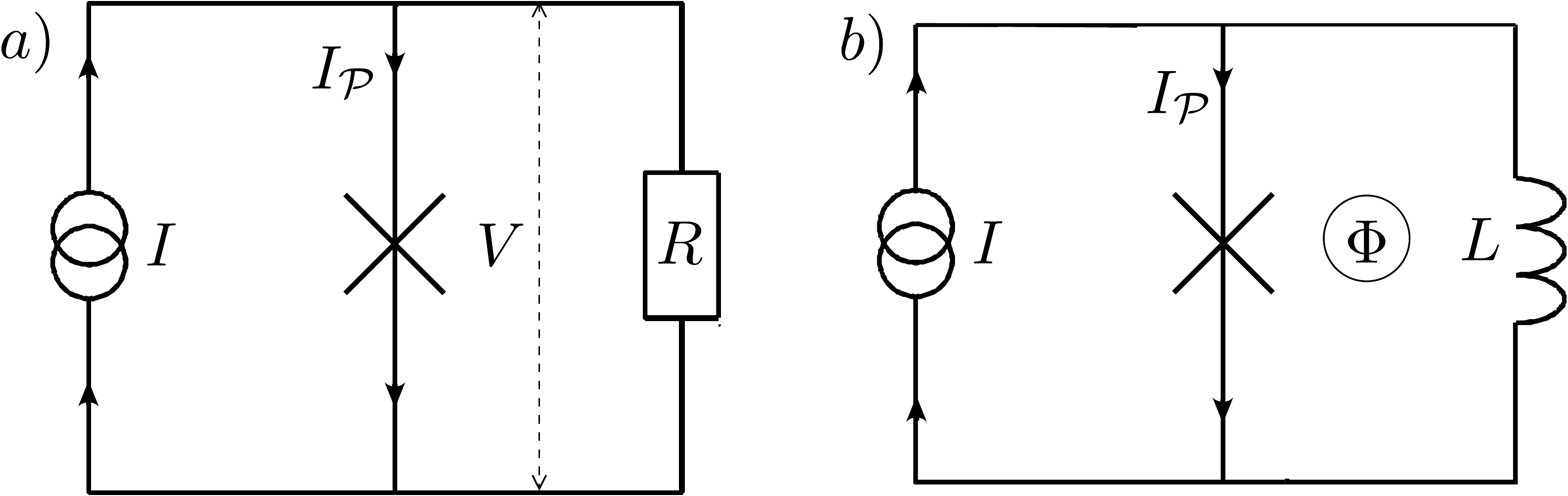}}
\caption{Left panel \textit{a}: Circuit of a current-biased, resistively shunted Josephson junction, to measure the current-voltage characteristic. Right panel \textit{b}: Circuit of an rf {\sc squid} to measure the current-phase relationship.}
\label{current-bias}
\end{figure}

We start by considering the current-biased circuit of Fig.~\ref{current-bias}a. A voltage $V=(\hbar/2e)d\phi/dt$ drops over a resistor $R$ in parallel with the Josephson junction, of capacitance $C$. The two characteristic time scales of the circuit are the $RC$ time and the phase relaxation time
\begin{equation}
\tau_{\rm J}=\frac{h}{2e}\frac{1}{RI_{4\pi,{\rm c}}}=\frac{R_{\rm Q}}{2R} \frac{\hbar}{E_{\rm T}},\label{tauJdef}
\end{equation}
where  $R_{\rm Q} = h/e^2$ is the resistance quantum. The characteristic energy scales of the Josephson junction are the charging energy $e^2/C$ and the Josephson energy $\hbar I_{4\pi,{\rm c}}/e=E_{\rm T}$.

The capacitance should be suffiently small that the phase dynamics is overdamped, $RC\ll\tau_{\rm J}$, and sufficiently large that the phase dynamics is classical, $e^{2}/C\ll E_{\rm T}$. We also wish to ensure that the Josephson junction remains in its ground state during the phase relaxation, which requires $E_{\rm T}\tau_{\rm J}/\hbar\gg 1$. Together these three conditions are met if
\begin{equation}
(R/R_{\rm Q})^{2}\tau_{\rm J}\ll RC\ll\tau_{\rm J},\label{tauJconditions}
\end{equation}
with $R\ll R_{\rm Q}$.

At a fixed parity ${\cal P}$, the bias current $I=I_{\cal P}+V/R$ drives the phase $\phi(t)$ according to
\begin{equation}
\frac{I}{I_{4\pi,{\rm c}}}=\frac{\tau_{\rm J}}{2\pi}\frac{d\phi}{dt}+\frac{1}{2\pi}{\rm mod}_{4\pi}(\phi+{\cal P}\pi+\pi)-1.\label{circuit-even}
\end{equation}
A typical value $E_{\text{T}}/k_{\rm B} \simeq 1\,{\rm K}$ gives $\hbar/E_T \simeq 10^{-11}\,{\rm s}$. The typical time scales for quasiparticle poisoning are in the $\mu{\rm s}$ to ms range \cite{Rai12}, so even if $R\ll R_{\rm Q}$ we can safely assume that $\tau_{\rm J} \ll \tau_{\text{qp}}$ and use Eq.\ \eqref{circuit-even} to calculate the relaxation of the phase in between two quasiparticle tunneling events.

The phase relaxation due to a quasiparticle tunneling event at $t=0$ (by which ${\cal P}\mapsto-{\cal P}$) amounts to a $2\pi$ phase slip on a time scale $\tau_{\rm J}$,
\begin{equation}
\phi(t)=\phi(0) e^{-t/\tau_{\rm J}}+[\phi(0)+2\pi] (1-e^{-t/\tau_{\rm J}}).\label{phirelax}
\end{equation}
Before and after the phase slip the junction is in a zero-voltage state, for bias currents $I\lesssim I_{4\pi,{\rm c}}$. During the phase slip there is a voltage pulse of integrated area $\int V(t)dt=h/2e$. The corresponding time-averaged voltage $\bar{V}=h/2e\tau_{\rm qp}$ is smaller by a factor $\tau_{\rm J}/\tau_{\rm qp}\ll 1$ than the voltage that develops for $I\gtrsim I_{4\pi,{\rm c}}$.

This shows that the current-biased circuit of Fig.~\ref{current-bias}a provides a {\sc dc} measurement of the parity-constrained critical  current $I_{4\pi, \text{c}}$. In contrast, the circuit of Fig.~\ref{current-bias}b is not sensitive to parity constraints. This rf {\sc squid} is phase-biased for sufficiently small inductance $L\ll h/eI_{4\pi,{\rm c}}$. Quasiparticle tunneling events have only a small effect on the phase, which remains fixed by the enclosed flux $\Phi=(\hbar/2e)\phi\approx LI$.

At low temperatures the parity of the number of electrons ${\cal N}$ in the system will equilibrate at the ground-state fermion parity $\sigma$, which implies that ${\cal N}$ is even (${\cal P}=1$) for ${\rm mod}_{4\pi}(\phi+\pi)<2\pi$ and odd (${\cal P}=-1$) for ${\rm mod}_{4\pi}(\phi+\pi)>2\pi$. In either case the supercurrent ${\cal I}_{\cal P}$ given by Eq.\ \eqref{current-phase} cannot become larger than $I_{4\pi,{\rm c}}/2=I_{2\pi,{\rm c}}$ --- which is the critical current without parity constraints.


\begin{thebibliography}{99}
\bibitem{Ma93} M. Ma and A. Yu. Zyuzin, Europhys. Lett. \textbf{21}, 941 (1993).
\bibitem{Ost11} J. A. M. van Ostaay, A. R. Akhmerov, and C. W. J. Beenakker, Phys. Rev. B \textbf{83}, 195441 (2011).
\bibitem{Sto11} M. Stone and Y. Lin, Phys. Rev. B \textbf{83}, 224501 (2011).
\bibitem{Ric12} P. Rickhaus, M. Weiss, L. Marot, and C. Sch\"{o}nenberger, Nano Lett. \textbf{12}, 1942 (2012).
\bibitem{Pop12} M. Popinciuc, V. E. Calado, X. L. Liu, A. R. Akhmerov, T. M. Klapwijk, and L. M. K. Vandersypen, Phys. Rev. B \textbf{85}, 205404 (2012).
\bibitem{Kom12} K. Komatsu, C. Li, S. Autier-Laurent, H. Bouchiat, and S. Gu\'{e}ron, Phys. Rev. B \textbf{86}, 115412 (2012).
\bibitem{Fu09} L. Fu and C. L. Kane, Phys. Rev. B \textbf{79}, 161408(R) (2009).
\bibitem{Kit01} A. Yu. Kitaev, Phys. Usp. \textbf{44} (suppl.), 131 (2001).
\bibitem{Kwo03} H.-J. Kwon, K. Sengupta, and V. M. Yakovenko, Braz. J. Phys. \textbf{33}, 653 (2003); Eur. Phys. J. B \textbf{37}, 349 (2004).
\bibitem{Gol04} A. A. Golubov, M. Y. Kupriyanov, and E. IlÕichev, Rev. Mod. Phys. \textbf{76}, 411 (2004).
\bibitem{Lut10} R. M. Lutchyn, J. D. Sau, and S. Das Sarma, Phys. Rev. Lett. \textbf{105}, 077001 (2010).
\bibitem{Law11} K. T. Law and P. A. Lee, Phys. Rev. B \textbf{84}, 081304 (2011).
\bibitem{Ios11} P. A. Ioselevich and M. V. Feigel'man, Phys. Rev. Lett. \textbf{106}, 077003 (2011).
\bibitem{Nog12} F. S. Nogueira and I. Eremin, J. Phys. Condens. Matter \textbf{24} 325701 (2012).
\bibitem{Bad11} D. M. Badiane, M. Houzet, and J. S. Meyer, Phys. Rev. Lett. \textbf{107}, 177002 (2011).
\bibitem{San12} P. San-Jose, E. Prada, and R. Aguado, Phys. Rev. Lett. \textbf{108}, 257001 (2012).
\bibitem{Dom12} F. Dom\'{\i}nguez, F. Hassler, and G. Platero, Phys. Rev. B \textbf{86}, 140503(R) (2012).
\bibitem{Pik11} D. I. Pikulin and Yu. V. Nazarov, Phys. Rev. B \textbf{86}, 140504(R) (2012).
\bibitem{Kon08} M. K\"{o}nig, H. Buhmann, L. W. Molenkamp, T. Hughes, C.-X. Liu, X.-L. Qi, and S.-C. Zhang, J. Phys. Soc. Jpn. \textbf{77}, 031007 (2008).
\bibitem{Bee91a} C. W. J. Beenakker and H. van Houten, Phys. Rev. Lett. \textbf{66}, 3056 (1991); extended version at arXiv:cond-mat/0512610.
\bibitem{note1} It is a pervasive misunderstanding that the factor $g=2$ in the relation \eqref{Iphirelation} between supercurrent and Andreev levels accounts for the charge $2e$ of the Cooper pairs, rather than counting the spin degeneracy. Because of this misunderstanding, the results for the supercurrent carried by Majorana zero-modes (which have $g=1$) should be divided by two in Ref.\ \onlinecite{Kwo03} and many follow-up papers. The origin of the misunderstanding is discussed in more detail by N. M. Chtchelkatchev and Yu.\ V. Nazarov, Phys. Rev. Lett. \textbf{90}, 226806 (2003).
\bibitem{Ish71} C. Ishii, Prog. Theor. Phys. (Kyoto) \textbf{44}, 1525 (1970).
\bibitem{Bar72} J. Bardeen and J. L. Johnson, Phys. Rev. B \textbf{5}, 72 (1972).
\bibitem{Svi73} A. V. Svidzinsky, T.  N. Antsygina, and E. N. BratusÕ, J. Low Temp. Phys. \textbf{10}, 131 (1973).
\bibitem{Bee91b} C. W. J. Beenakker, Phys. Rev. Lett. \textbf{67}, 3836 (1991); extended version at arXiv:cond-mat/0406127.
\bibitem{Tuo92} M. T. Tuominen, J. M. Hergenrother, T. S. Tighe, and M. Tinkham, Phys. Rev. Lett. \textbf{69}, 1997 (1992).
\bibitem{note2} The factor of two in the relation \eqref{IpmZpm} between supercurrent and free energy accounts for the Cooper pair charge, with all degeneracies included in the partition function. Please see Ref.\ \onlinecite{note1} to avoid any misunderstanding.
\bibitem{Bro97} P. W. Brouwer and C. W. J. Beenakker, Chaos, Solitons \& Fractals \textbf{8}, 1249 (1997). Several misprints are corrected in the online version at arXiv:cond-mat/9611162.
\bibitem{App} Details of the calculations are given in the Appendices.
\bibitem{note3} The basis chosen for the Bogoliubov-De Gennes Hamiltonian \eqref{HBdGQSH} is \{spin-up electron, spin-down electron, spin-up hole, spin-down hole\}. In this basis the electron and hole blocks of $H_{\rm BdG}$ are minus each others complex conjugate and the electron and hole scattering matrices $s_{ee},s_{hh}$ are related by $s^{\ast}_{hh}(-\varepsilon)=s_{ee}(\varepsilon)\equiv s_{0}(\varepsilon)$. Similarly, the Andreev reflection matrices $s_{he},s_{eh}$ from electron to hole and from hole to electron are related by $s^{\ast}_{eh}(-\varepsilon)=s_{he}(\varepsilon)\equiv r_{\rm A}(\varepsilon)$.
\bibitem{Kon07} M. K\"{o}nig, S. Wiedmann, C. Br\"{u}ne, A. Roth, H. Buhmann, L. W. Molenkamp, X.-L. Qi, and S.-C. Zhang, Science \textbf{318}, 766 (2007).
\bibitem{Kne11a} I. Knez, R.-R. Du, and G. Sullivan, Phys. Rev. Lett. \textbf{107}, 136603 (2011).
\bibitem{Kne11b} I. Knez, R.-R. Du, and G. Sullivan, Phys. Rev. Lett. \textbf{109}, 186603 (2012).
\bibitem{Rok12} L. P. Rokhinson, X. Liu, and J. K. Furdyna, Nature Phys. \textbf{8}, 795 (2012) .
\bibitem{Rai12} D. Rainis and D. Loss, Phys. Rev. B \textbf{85}, 174533 (2012).
\bibitem{Mou12} V. Mourik, K. Zuo, S. M. Frolov, S. R. Plissard, E. P. A. M. Bakkers, and L. P. Kouwenhoven, Science \textbf{336}, 1003 (2012).
\bibitem{Den12} M. T. Deng, C. L. Yu, G. Y. Huang, M. Larsson, P. Caroff, and H. Q. Xu, Nano Lett. \textbf{12}, 6414 (2012).
\bibitem{Das12} A. Das, Y. Ronen, Y. Most, Y. Oreg, M. Heiblum, and H. Shtrikman, Nature Phys. \textbf{8}, 887 (2012).
% references from appendix
\bibitem{Ryu10} S. Ryu, A. Schnyder, A. Furusaki, and A. Ludwig, New J. Phys. \textbf{12}, 065010 (2010).
\bibitem{Akh11} A. R. Akhmerov, J. P. Dahlhaus, F. Hassler, M. Wimmer, and C.W.J. Beenakker, Phys. Rev. Lett. \textbf{106}, 057001 (2011).
\bibitem{Die12} M. Diez, J. P. Dahlhaus, M. Wimmer, and C. W. J. Beenakker, Phys. Rev. B \textbf{86}, 094501 (2012).
\end{thebibliography}
\end{document}